\documentclass{article}

\usepackage[english]{babel}

\usepackage[letterpaper,top=2cm,bottom=2cm,left=3cm,right=3cm,marginparwidth=1.75cm]{geometry}

\usepackage{amsmath}
\usepackage{amssymb}
\usepackage{graphicx}
\usepackage[colorlinks=true, allcolors=blue]{hyperref}
\usepackage{url}
\usepackage{caption}
\usepackage{subcaption}

\title{The Floyd-Warshall Algorithm Re-implemented Using 3D-Tensors and Hardware Acceleration}
\author{Taher Anjary}

\begin{document}
\maketitle

\begin{abstract}
The Floyd-Warshall(FW) algorithm\cite{Cormen1990IntroductionTA}, is an ancient but a largely important algorithm used to solve the all-pairs simple-paths(APSP) problem. While the algorithm is available for use in open-source graph optimization libraries such as NetworkX\cite{SciPyProceedings_11}, they do not take advantage of modern parallel processing hardware such as Graphics Processing Units(GPUs), which would reduce compute time to a fraction of its iterative or recursive implementations. In this work, a re-implementation of the Floyd-Warshall algorithm using open-source GPU libraries such as PyTorch\cite{paszke2017automatic} is presented. A further implementation of the R-Kleene\cite{DAlberto2006RKleeneAH} is also described, a slightly newer algorithm used for solving the APSP problem in a divide-and-conquer, recursive but highly parallelized architecture. In addition, a random graph generator that generates a wide range of graphs of different scales is also contributed, where the densities and connectivities are controlled using some heuristics. The run-times of the GPU accelerated FW algorithm and R-Kleene on these heuristically generated graphs are evaluated against each other and to the widely used implementation from NetworkX. The code for the GPU implementation of the algorithms, the random graph generator, and the Blender animation file are available at \url{https://github.com/tanjary21/APSP_GPU/}.
\end{abstract}

\section{Introduction}

The All-Pairs Shortest Path(APSP) problem requires that we determine shortest path distances between every pair of nodes in a network. One approach to solving this problem is via a general \textit{all-pairs label-correcting algorithm}. Such algorithms are guaranteed to yield a solution for dense networks hence they generalize to sparse networks as well. One such example of a popular algorithm is the \textit{Floyd-Warshall} algorithm, which runs in $O(n^3)$ time. \\

The most freely available implementation of the \textit{Floyd-Warshall} algorithm is from NetworkX\cite{SciPyProceedings_11}, a popular network analysis library in Python that allows academics and industrialists to perform creation, manipulation, and study of the structure, dynamics, and functions of complex networks. However, the package is restricted for use on Central-Processing-Units(CPU) machines, making it unscalable for larger graphs in large domains beyond logistics, space travel and internet routing where real-time performance may be demanded. Modern Graphics-Processing-Units(GPUs) are an excellent powerhouse for processing power and computation. Their intended use is in the gaming industry but they have recently been adopted in academic research to perform chemical simulations, protein folding and train deep neural networks in the artificial intelligence domain. This is only possible due to the extremely parallelized Tensor cores made available through hardware acceleration. This specifically means that large matrix operations on hyper-dimensional matrices(termed as Tensors) are performed in a fraction of the time that it would take a CPU to do the same.\\

In this work, I conjecture that the APSP problem can be augmented into a clever matrix, and a custom matrix operation kernel known as \textit{min-plus} can be implemented in a similar way to matrix-matrix multiplication. This then allows the APSP problem to be processed on GPUs that will yield a result in a fraction of the time of NetworX's CPU implementation. Further, GPU-friendly APSP algorithms such as \textit{R-Kleene} is also explored and benchmarked. I show that the GPU implementation of Floyd-Warshall runs significantly faster than the CPU implementation, and further show that R-Kleene is even faster at scale. Graphs of a wide variety of scales and densities are used for the benchmarking and show computational complexity trends. Furthermore, the implementations are made publicly available at \url{https://github.com/tanjary21/APSP_GPU/} and is implemented in Pytorch\cite{paszke2017automatic}, an open-source, automatic-differentiation framework freely available to all and runs in Python environments.

\section{Background: The Floyd-Warshall Algorithm}

"Given a matrix of distances $d[i,j]$ between a source node $i$ and target node $j$, the \textit{Floyd-Warshall} algorithm maintains a shortest path distances within $O(n^3)$ computations. The algorithm achieves this bound by applying the triple operations cleverly. The algorithm is based on inductive arguments developed by an application of a dynamic programming technique.\\

Let $d^k[i, j]$ represent the length of a shortest path from node $i$ to node $j$ subject
to the condition that this path uses only the nodes 1, 2, ... ,k - 1 as internal nodes.
Clearly, $d^{n + 1} [i, j]$ represents the actual shortest path distance from node $i$ to node
$j$. The Floyd-Warshall algorithm first computes $d^1[i, j]$ for all node pairs $i$ and $j$. Using $d^1[i, j]$, it then computes $d^2[i, j]$ for all node pairs $i$ and $j$. It repeats this process until it obtains $d^{n + 1} [i, j]$ for all node pairs $i$ and $j$, when it terminates. Given $d^k[i, j]$, the algorithm computes $d^{k + 1} [i, j]$ using the following property:
\begin{equation}
    d^{k+1}[i,j] = min\{ d^k[i, j], d^k[i, k] + d^k[k, j] \}
\end{equation}

This property is valid for the following reason. A shortest path that uses only
the nodes 1,2, ... , k as internal nodes either (1) does not pass through node $k$, in
which case $d^{k + 1} [i, j] = d^k[i, j]$, or (2) does pass through node $k$, in which case $d^{k + 1} [i, j] = d^k[i, k] + d^k[k, j]$. Therefore, $d^{k+1}[i,j] = min\{ d^k[i, j], d^k[i, k] + d^k[k, j] \}$.\\

The Floyd-Warshall algorithm uses predecessor indices, $pred[i, j]$, for each
node pair $[i, j]$. The index $pred[i, j]$ denotes the last node prior to node $j$ in the
tentative shortest path from node $i$ to node $j$. The algorithm maintains the invariant
property that when $d[i, j]$ is finite, the network contains a path from node $i$ to node
$j$ of length $d[i, j]$. Using the predecessor indices, we can obtain this path, say
$P$, from node $k$ to node $l$ as follows. We backtrack along the path $P$ starting at
node $l$. Let $g = pred[k, l]$. Then $g$ is the node prior to node $l$ in $P$. Similarly, $h =
pred[k, g]$ is the node prior to node $g$ in $P$, and so on. We repeat this process until
we reach node $k$.\\

The Floyd-Warshall algorithm clearly performs $n$ major iterations, one for each
$k$, and within each major iteration, it performs $0(1)$ computations for each node pair.
Consequently, it runs in $0(n^3)$ time\cite{Cormen1990IntroductionTA}."

\section{Methodology}

The ability to concatenate paths can be delegated to matrix operations.\\

Suppose $X \in \mathbb{R}^{N_1 \times N_2}$ whose rows represent a set $R$ of $N_1$ vertices and whose columns represent a set $S$ of $N_2$ vertices. Let $X_{i,k}$ retrieve the cost of going from a vertex $i \in R$ to a vertex $k \in S$. $R$ and $S$ can be the same set of vertices but is not necessary. Let $Y \in \mathbb{R}^{N_2 \times N_3}$ be another cost matrix such that the entries at $Y_{k,j}$ represent the cost of going from a vertex $k \in S$ to a vertex $j \in T$. $T$ can be the same as $R$ and $S$ but is also not necessary. Then, the minimum cost $Z_{i,j}$ in $Z \in \mathbb{R}^{N_1 \times N_3}$ of going from a vertex $i \in R$ to a vertex $j \in T$ via an intermediate vertex $k \in S$ can be computed via the following expression:
\begin{equation}
\label{minplus}
     Z_{i,j} = min( X_{i,1}+Y_{1,j}, X_{i,2}+Y_{2,j}, ... , X_{i,N_2}+Y_{N_2,j} )
\end{equation}

If we observe equation \ref{minplus} closely, we can observe that it is similar to matrix multiplication:
\begin{equation}
\label{mm}
     Z_{i,j} = sum( X_{i,1} \times Y_{1,j}, X_{i,2} \times Y_{2,j}, ... , X_{i,N_2} \times Y_{N_2,j} )
\end{equation}

i.e. do an element wise multiplication for row $i$ in $X$ and each column $j$ in $Y$, sum the products, and insert the result into $Z$ at the $i^{th}$ row and $j^{th}$ column. If we instead replace the \textit{sum} in \ref{mm} with \textit{min}, and the \textit{$\times$} with \textit{+}, we get equation \ref{minplus}\cite{CUDAAPSP}.\\

The main source of efficiency comes from the ability to do matrix operations on highly optimized parallel computation using hardware acceleration devices such as GPUs. This means that we can do the operation described by equation \ref{minplus} in one shot, for all $k \in N_2$, without for-loops, for all $i,j \in N_1, N_3$, by simply writing a kernel that replaces matrix multiplication in equation \ref{mm} with \textit{min-plus}, as described by equation \ref{minplus}.\\

\subsection{Min-Plus Operation on GPU}

I implement \textit{min-plus} by utilizing the \textit{broadcast addition}, \textit{minimum operation} and \textit{argmin operation} on \textit{3D Tensors}. Each of these operations are highly parallelized and available for free usage from PyTorch\cite{paszke2017automatic}, an open-source framework for hardware accelerated, parallel computation.\\

Suppose we have our original cost matrix $H \in \mathbb{R}^{N \times N \times 1}$, which represents the cost of going from any vertex $i$, indexed by the row, to any other vertex $j$, indexed by the column. Let there be no negative edge-cycles, and no edge with 0 cost, except for self-loops. If we add $H$ to its 3D transpose $H' \in \mathbb{R}^{1 \times N \times N}$ via \textit{broadcast addition}, we get a 3D Tensor $L \in \mathbb{R}^{N \times N \times N}$. The first dimension in $L$ indexes a path starting \textit{from} a vertex $i$. The second indexes a path travelling via an \textit{intermediate} vertex $k$. The third indexes a path that ends at a \textit{to} vertex $j$. Each element $L_{i,k,j} \in L$ hence represents, exhaustively for every $i,k,j \in N, N, N$, the concatenated cost of going from any vertex $i$ to any intermediate vertex $k$ and from intermediate vertex $k$ to vertex $j$. If we then perform a \textit{minimum} operation along the second dimension in $L$, we collapse $L$ to a new, updated cost matrix $H_{new} \in \mathbb{R}^{N \times 1 \times N}$, which we can 3D-transpose to have the shape $\mathbb{R}^{N \times N \times 1}$. $H_{new}$ will now be updated with shortest paths that are at most 1-hop-neighbors apart. If we then perform a \textit{argmin} operation along the second dimension in $L$, we obtain the \textit{indices} of the minimum operation, the result is matrix $P \in \mathbb{R}^{N \times 1 \times N}$, which corresponds to the updated predecessors of the shortest paths found so far.

\begin{figure}[!htb]
\minipage{0.137142857\textwidth}
  \includegraphics[width=\linewidth]{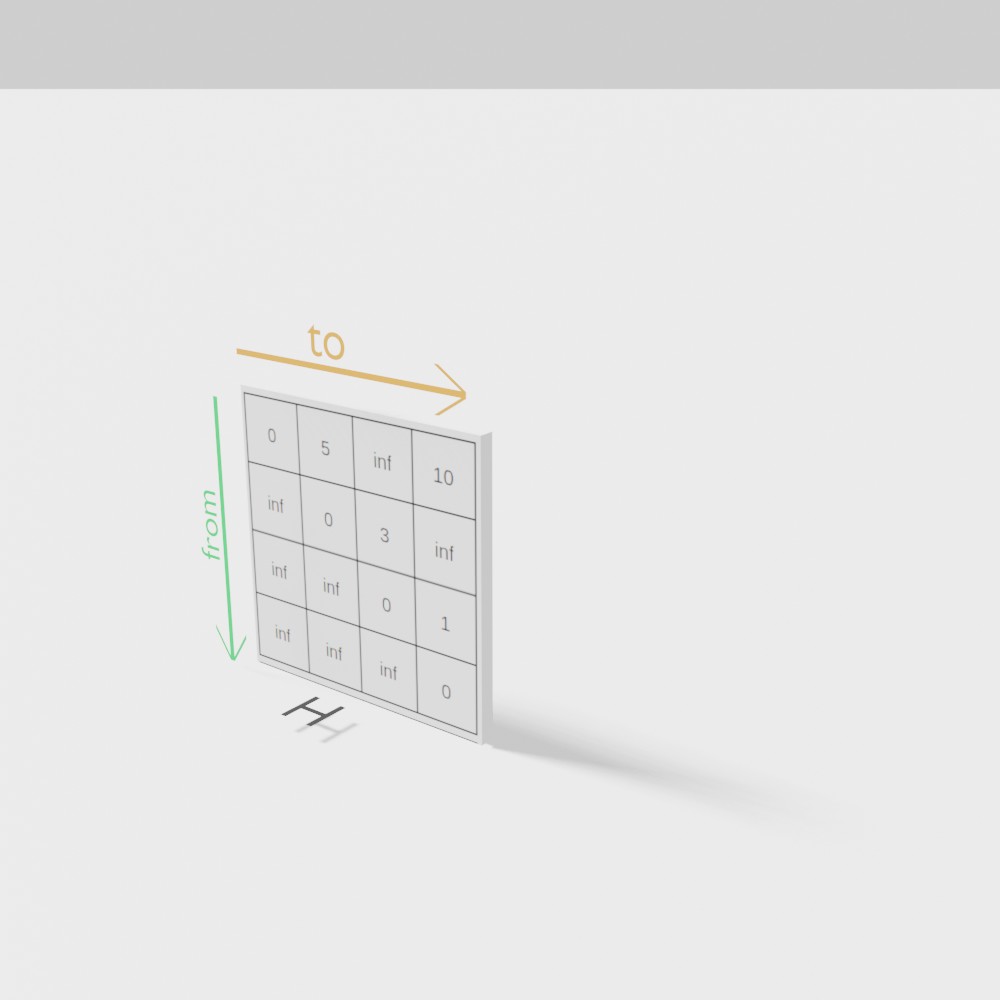}
  \caption{Step 0}
\endminipage\hfill
\minipage{0.137142857\textwidth}
  \includegraphics[width=\linewidth]{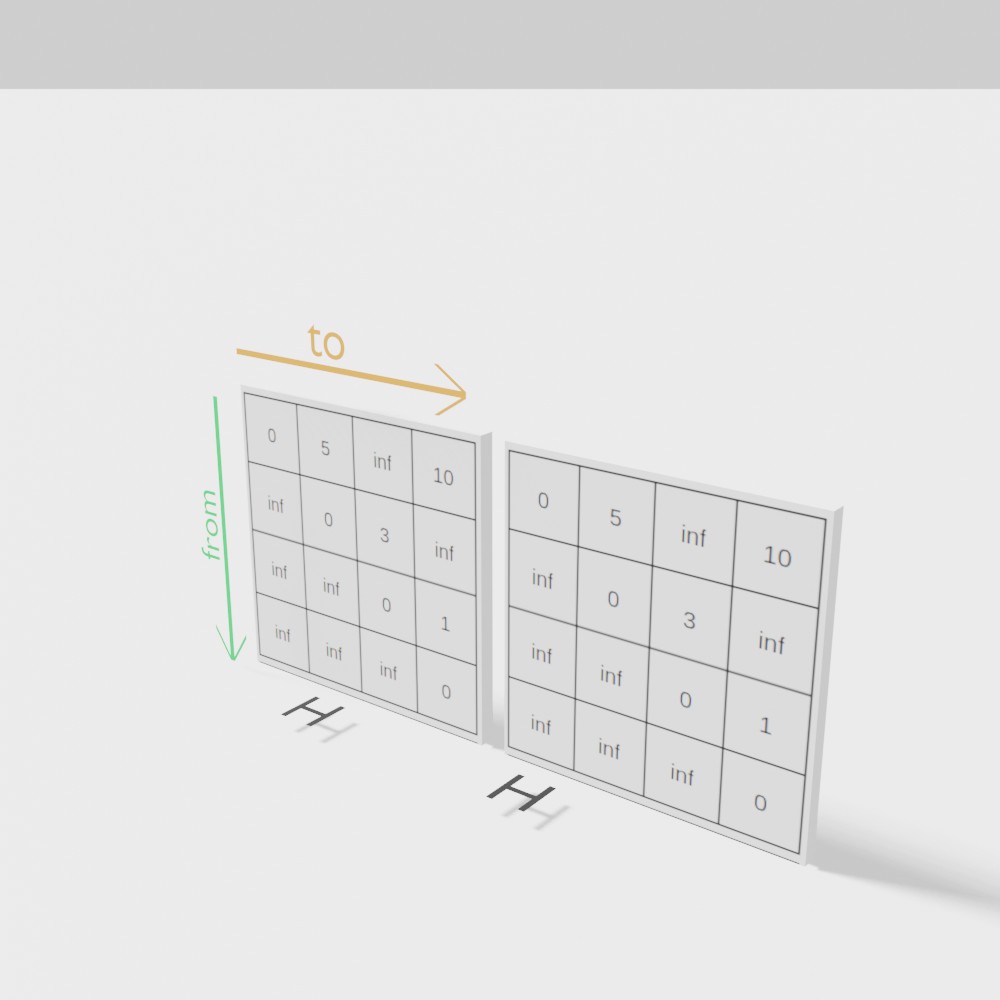}
  \caption{Step 1}
\endminipage\hfill
\minipage{0.137142857\textwidth}
  \includegraphics[width=\linewidth]{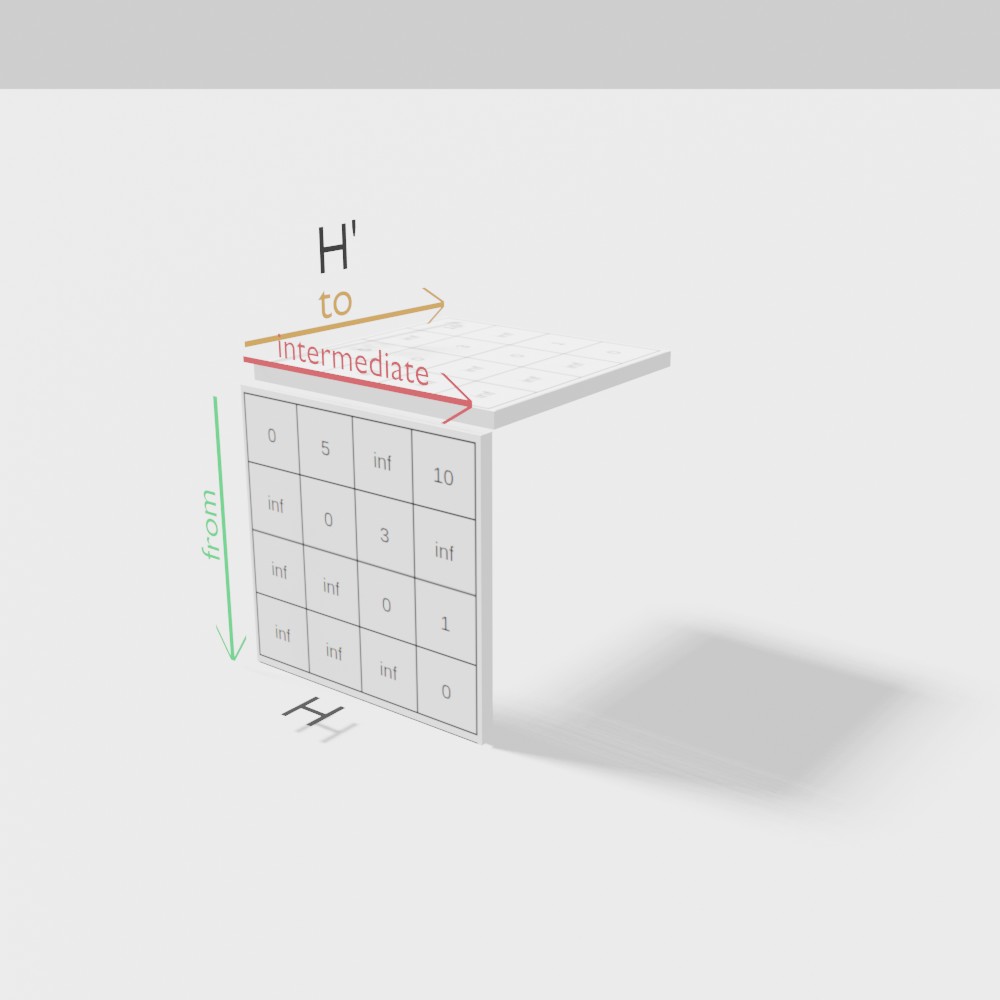}
  \caption{Step 2}
\endminipage\hfill
\minipage{0.137142857\textwidth}
  \includegraphics[width=\linewidth]{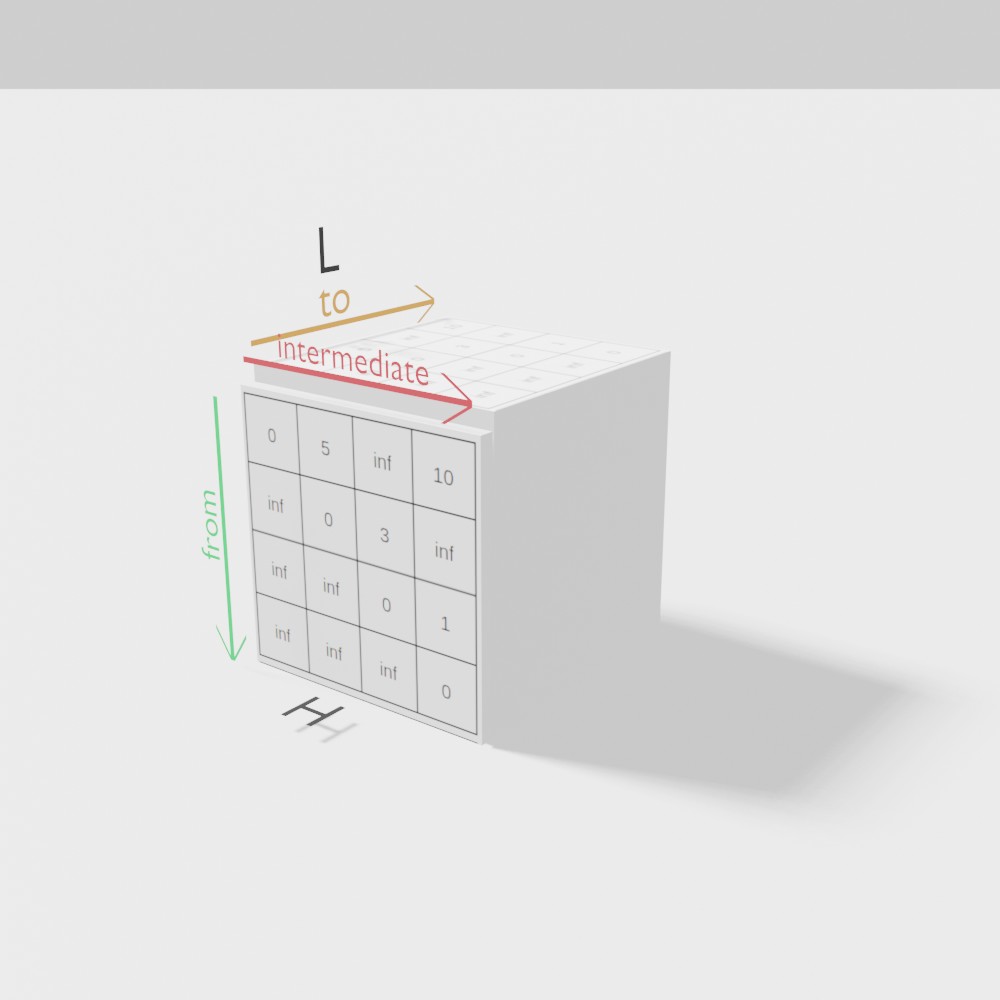}
  \caption{Step 3}
\endminipage\hfill
\minipage{0.137142857\textwidth}
  \includegraphics[width=\linewidth]{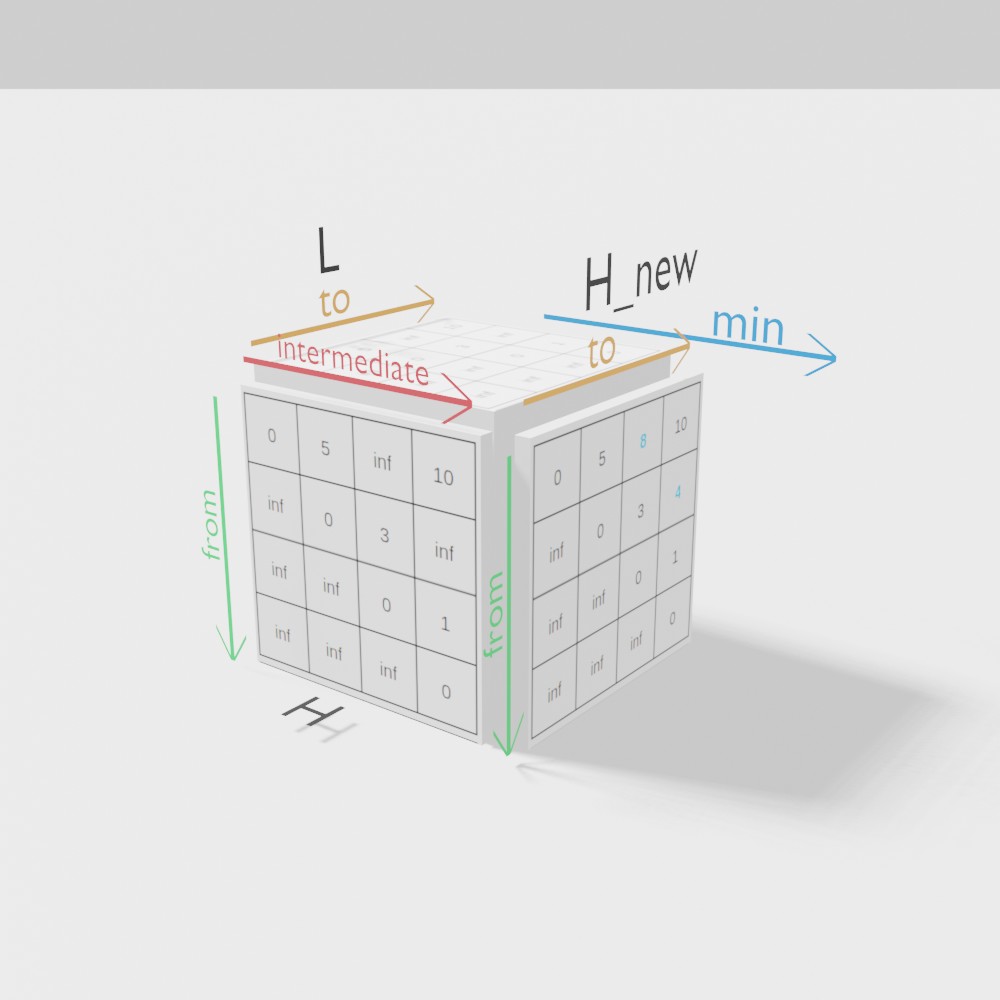}
  \caption{Step 4}
\endminipage\hfill
\minipage{0.137142857\textwidth}
  \includegraphics[width=\linewidth]{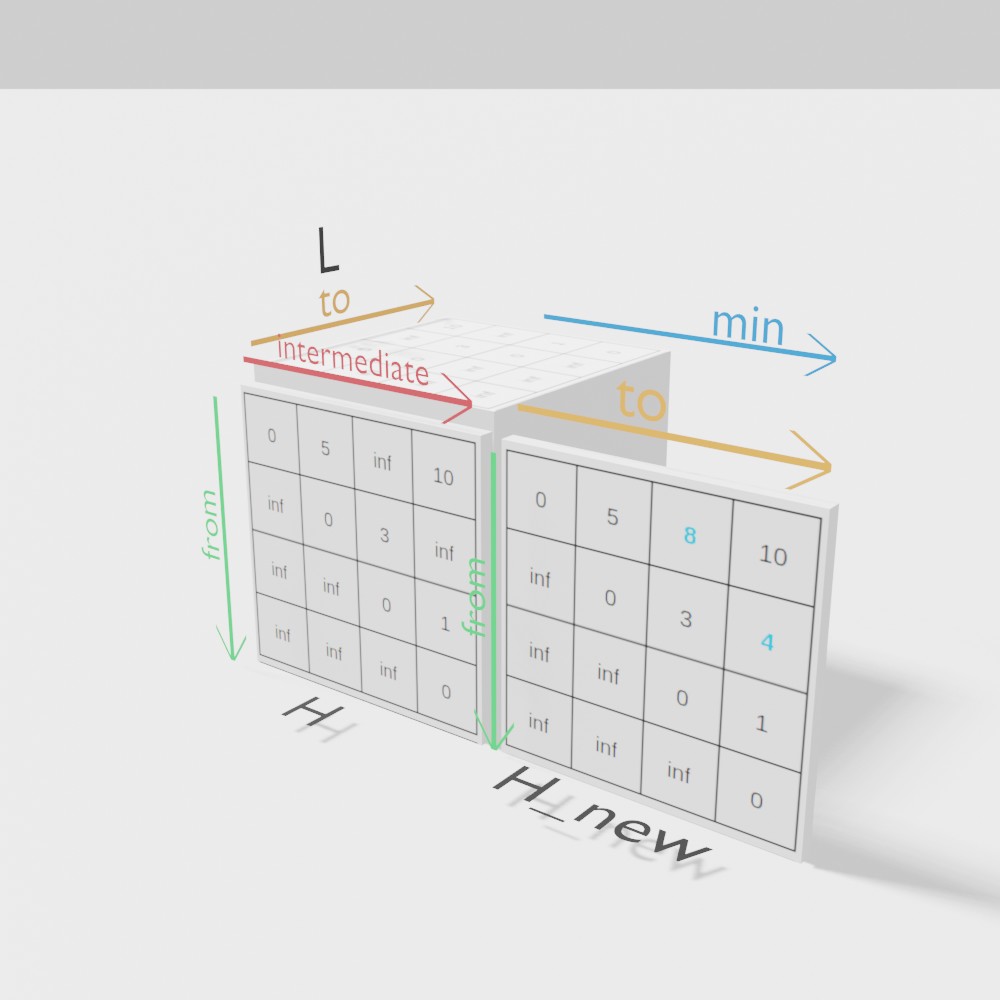}
  \caption{Step 5}
\endminipage\hfill
\minipage{0.137142857\textwidth}%
  \includegraphics[width=\linewidth]{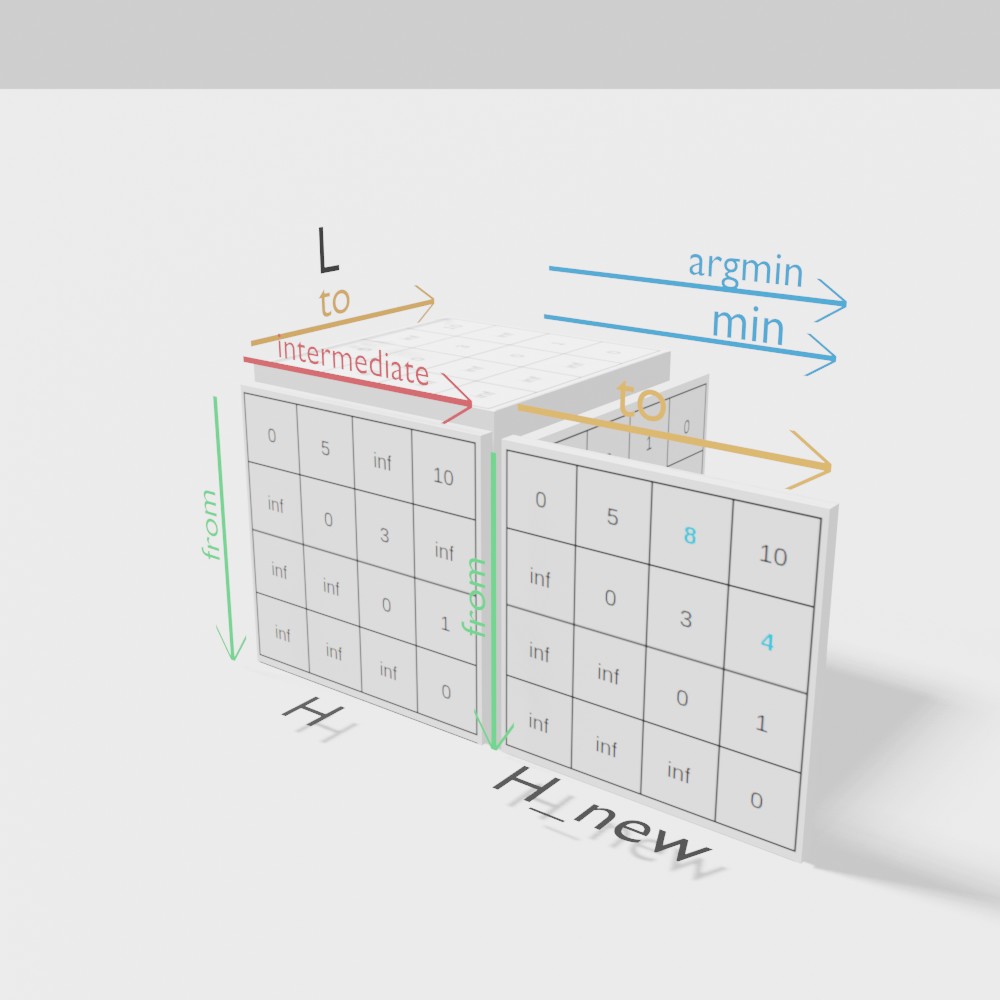}
  \caption{Step 6}
\endminipage
\caption{Step 0: Initial cost matrix H, Step 1: Clone H', Step 2: 3D Transpose H', Step 3: Generate 3D Tensor L via broadcast addition, Step 4: Perform min on dimension 2, Step 5: Updated cost matrix $H_{new}$, Step 6: Perform argmin on dimension 2 gives predecessors}
\end{figure}

\subsection{Floyd-Warshall on GPU}
By simply running the \textit{min-plus} operation repeatedly until we observe no changes between $H^{t}_{new}$ and $H^{t+1}_{new}$, we will have implemented the Floyd-Warshall algorithm with hardware acceleration on the GPU. This outer loop is bounded by $N$, since we may have shortest paths that are in the worst case N-hop-neighbors apart.

\subsection{R-Kleene on GPU}
The R-Kleene algorithm\cite{DAlberto2006RKleeneAH} for solving the APSP problem also utilizes the \textit{min-plus} operation, but it does so by calling it recursively. It works by first chunking the original cost matrix $H$ into 4 blocks.
\begin{equation}
H =
\begin{bmatrix}
A & B\\
C & D
\end{bmatrix}
\end{equation}

Block $A$ has costs of edges between vertices in the first half of the graph. Block $D$ has costs of edges between vertices in the second half. Block $B$ has costs of edges from the first half to the second half, and block $C$ has costs of edges back from the second half to the first half.\\

R-Kleene first calls itself recursively on $A$, to get shortest paths that never leave the first half of the graph. Then, by using \textit{min-plus} it updates $B$, $C$, and $D$ to account for paths that go through $A$. Then it makes another recursive call on $D$, to find shortest paths from one vertex in $D$ to another (which now includes paths that go through $A$ on the way). Then, using \textit{min-plus}, it again updates $B$ and $C$ and finally $A$ to account for those paths too\cite{CUDAAPSP}.\\

This divide-and-conquer approach implies a logarithmic time complexity.

\subsection{Graph Generator}
In order to benchmark the performance of hardware accelerated APSP implementations against NetworkX's, a random graph generator was written to generate a broad range of graphs of varying nodes, and edge densities(connectivity).

The graph generator can be represented as a the function $G = f(V, \rho, \alpha)$ where $V$ indicates the desired number of nodes, $\rho$ determines the density of edges and $\alpha$ scales the costs of each edge. The algorithm starts by sampling a probability matrix $P \in \mathbb{R}^{V \times V}$. By multiplying $P$ with $\rho$, we scale the probabilities' likelihoods of becoming a 1. This matrix is then passed to an independent Bernoulli process that converts the probability matrix $\rho P$ to an adjacency matrix $A \in \{0,1\}  \mathbb{R}^{V \times V}$. By then assigning random costs to each edge in $A$ by sampling random integers from the range [0, $\alpha$], we generate the cost matrix $H$. Note that the diagonal in $H$ are assigned zeros to give self-loops a zero cost. $H$ is then used as the input to all the APSP algorithms. I used $\alpha=100$ and $\rho$ a random number chosen uniformly from $[0, 100]$. I generate 1000 graphs using the generator $f$, each generated randomly with $V$(number of nodes) also being a random number chosen uniformly between $[4, 1000]$

\section{Experiments}

\begin{figure}[h]
\centering
    \begin{subfigure}[b]{0.49\textwidth}
    \centering
    \includegraphics[width=\textwidth]{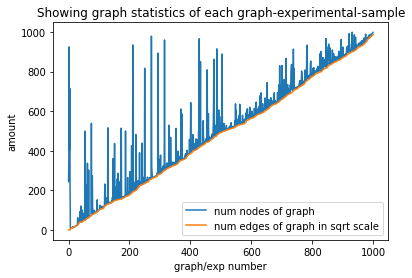}
    \caption{Graph statistics}
    \label{graph_stats}
    \end{subfigure}
    \hfill
    \begin{subfigure}[b]{0.49\textwidth}
    \centering
    \includegraphics[width=\textwidth]{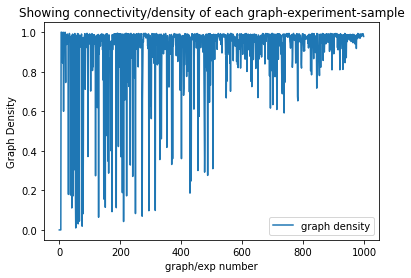}
    \caption{Graph densities}
    \label{graph_dens}
    \end{subfigure}
        \caption{Showing details of the generated graphs. (a) shows the square root of the number of edges in the graph and the number of nodes. The larger the gap between number of nodes and edges, the lower the density/connectivity, which is reflected in (b). The lower the gap, the higher the density. This is because the maximum number of edges that a graph of n nodes can have is $0.5 \times n(n-1)$, and this is reflected as a density of 1.0 in (b)}
        \label{graph_deets}
\end{figure}

1000 random graphs were generated using the graph generator. The graphs were then  sorted in ascending order of the number of edges in the graph. The statistics of the graphs are shown in Figure \ref{graph_deets}.

\begin{figure}[h]
\centering
    \begin{subfigure}[b]{0.49\textwidth}
    \centering
    \includegraphics[width=\textwidth]{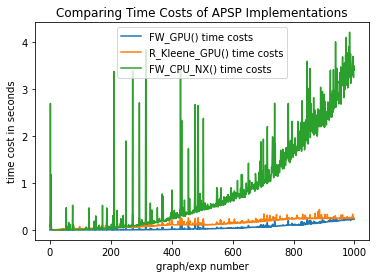}
    \caption{Time costs of each algorithm on each graph}
    \label{nx_cc}
    \end{subfigure}
    \hfill
    \begin{subfigure}[b]{0.49\textwidth}
    \centering
    \includegraphics[width=\textwidth]{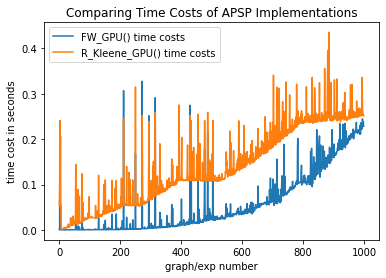}
    \caption{Time costs of only the GPU algorithms on each graph}
    \label{gpu_cc}
    \end{subfigure}
        \caption{Showing time costs of the three APSP alorithms: Floyd-Warshall on the GPU(FW-GPU), R-Kleene on the GPU(R-Kleene-GPU) and NetworkX's CPU implementation of Floyd-Wasrhall(FW-CPU-NX). (a) shows all three. (b) shows the same data as (a), but without NetworkX's time costs for better visibility.}
        \label{ccs}
\end{figure}

Then, I ran the 3 APSP implementations(FW-GPU, R-KLeene-GPU and Network X's FW-CPU-NX) on each of the graphs. I also record the time it takes for each algorithm to yield a solution for each graph. The records are shown in Figure \ref{ccs}. As can be observed from Figure \ref{nx_cc}, NetworkX's implementation is significantly slower than GPU versions and does not scale well to larger graphs. In ordre to compare the GPU implementations to each other, observe Figure \ref{gpu_cc}. It becomes clear that while using GPUs is faster than CPUs, clever algorithms are still necessary in order to be feasible at larger scales. What we see is that the FW-GPU is still polynomial but R-Kleene  is somewhat logarithmic. R-Kleene scales well and is more efficient for graphs with more than 1000 nodes. Unfortunately, I could not run tests with graphs of more than 1000 nodes due to memory limitations. I used an NVIDIA RTX 3090 GPU with 24 Gb of VRAM.

\section{Conclusion}

In this work, I implemented the Floyd-Warshall algorithm for the All-Pairs-Shortest-Paths(APSP) problem that exploit the matrix-matrix, hardware-accelerated, highly parallelized abilities of the GPU. I show that doing this is signifcantly faster than using widely avaialable implementations that run on CPUs. I further show that even though the GPU implementation is faster, it does not scale well. Therefore, divide-and-conquer algorithms such as R-Kleene, also implemented on the GPU, prevail at scale. I show that R-Kleene has logarithmic time-costs in a data-centric way rather than a theoretical way. By using GPUs, we alleviate the $O(n^3)$ runtimes, however, they end up consuming $n^3$ memory instead, which is why I could not run experiments for graphs larger than 1000 nodes. Ways to work around this is possible by attacking the memory demands in a divide-and-conquer way, similar to R-Kleene- in order to do this, we would need to divide the 3D-Tensor L. Other implementation-centric directions involve making use of multiple-GPUs. These are interesting directions to consider but are left for future work.

\bibliographystyle{alpha}
\bibliography{sample}

\end{document}